\documentclass[runningheads]{llncs}
\usepackage[T1]{fontenc}
\usepackage{graphicx}
\usepackage{subcaption}
\usepackage{amsmath, amssymb} 
\usepackage{tcolorbox}
\usepackage{xcolor}
\usepackage{nicematrix}
\usepackage{booktabs}
\usepackage{multirow}
\usepackage{graphicx}
\usepackage{makecell}
\usepackage{array}
\usepackage{siunitx}
\usepackage[font=scriptsize]{caption}
\usepackage{soul}

\newtcbox{\highlightedtext}{on line, boxrule=0pt, colback=yellow, colframe=yellow, sharp corners, boxsep=0pt, leftrule=0pt, rightrule=0pt, toprule=0pt, bottomrule=0pt}

\begin{document}
\title{Self-supervised learning for radio-astronomy source classification: a benchmark}
\titlerunning{SSL benchmark for radio-astronomy}

\author{
Thomas Cecconello   \inst{1,2}  \and
Simone Riggi        \inst{2}    \and
Ugo Becciani        \inst{2}    \and
Fabio Vitello       \inst{2}    \and
Andrew M. Hopkins   \inst{4}    \and 
Giuseppe Vizzari    \inst{3}    \and
Concetto Spampinato \inst{1}    \and 
Simone Palazzo      \inst{1}
}
\authorrunning{Cecconello et al.}
%
\institute{
Pattern Recognition and Computer Vision (PeRCeiVe) Lab, Department of Electrical, Electronics and Computer Engineering, University of Catania, Italy
\and
INAF - Osservatorio Astrofisico di Catania, Italy
\and
University of Milano Bicocca, Italy
\and
School of Mathematical and Physical Sciences, Macquarie University, Australia
}
\maketitle              
\begin{abstract}
The upcoming Square Kilometer Array (SKA) telescope marks a significant step forward in radio astronomy, presenting new opportunities and challenges for data analysis. Traditional visual models pretrained on optical photography images may not perform optimally on radio interferometry images, which have distinct visual characteristics.

Self-Supervised Learning (SSL) offers a promising approach to address this issue, leveraging the abundant unlabeled data in radio astronomy to train neural networks that learn useful representations from radio images. This study explores the application of SSL to radio astronomy, comparing the performance of SSL-trained models with that of traditional models pretrained on natural images, evaluating the importance of data curation for SSL, and assessing the potential benefits of self-supervision to different domain-specific radio astronomy datasets.

Our results indicate that, SSL-trained models achieve significant improvements over the baseline in several downstream tasks, especially in the linear evaluation setting; when the entire backbone is fine-tuned, the benefits of SSL are less evident but still outperform pretraining. These findings suggest that SSL can play a valuable role in efficiently enhancing the analysis of radio astronomical data. The trained models and code is available at: \url{https://github.com/dr4thmos/solo-learn-radio}
\end{abstract}

\section{Introduction}
Radio astronomy, a branch of astronomy that studies celestial objects through their radio emissions, has revolutionized our understanding of the universe. Unlike traditional optical telescopes, radio telescopes are essentially highly sensitive antennas designed to detect faint radio signals from space. These sophisticated instruments can range from single dish antennas to vast arrays of interconnected antennas spread over large distances. By capturing and analyzing these radio waves, astronomers can observe phenomena invisible to optical telescopes, penetrating cosmic dust and gas to reveal hidden aspects of our universe. This field is now on the cusp of a data revolution, with next-generation telescope arrays like the Square Kilometre Array (SKA) \cite{dewdney2016ska1}  set to generate unprecedented volumes of high-resolution data.

The SKA, an international effort to build the world's largest radio telescope, promises unparalleled sensitivity and survey speed. Its precursors, such as MeerKAT \cite{goedhart2023sarao} in South Africa and ASKAP \cite{norris2011emu} in Australia, are already producing vast amounts of high-quality data, foreshadowing the data deluge expected from SKA. This surge in data quantity and quality presents both opportunities and challenges for machine learning applications in astronomy.

Machine Learning (ML) techniques have become increasingly crucial in analyzing radio astronomical data. From source detection \cite{riggi2023astronomical} to classification \cite{riggi2024classification} and anomaly detection \cite{LOCHNER2021100481astronomaly}, ML algorithms are helping astronomers sift through terabytes of data efficiently. However, the unique characteristics of radio interferometry images pose challenges for traditional computer vision models, often pre-trained on optical images.

Self-Supervised Learning (SSL) \cite{9462394surveyssl} has emerged as a powerful paradigm to address these challenges. By leveraging large amounts of unlabeled data, SSL enables models to learn meaningful representations without manual annotations. This is particularly valuable in radio astronomy, where labeled datasets are often limited but unlabeled data is abundant. Moreover, the labeling schemes in radio astronomical datasets can vary significantly depending on the specific study or survey objectives, making it challenging to create large, consistently labeled datasets. SSL offers a way to leverage the vast amounts of unlabeled data while potentially bridging the gaps between different labeling conventions.

While recent works have explored SSL in radio astronomy \cite{huertas2023brief}, they often focus on a single SSL method or a limited set of downstream tasks. This leaves a gap in our understanding of how different SSL techniques perform across various radio astronomy datasets and tasks.

Our study aims to provide a comprehensive benchmark of SSL methods applied to radio astronomical images, with the following objectives:
\begin{itemize}
\item Evaluate the performance of SSL-trained models compared to traditional models pretrained on natural images across various radio astronomy tasks.
\item Assess the impact of data curation on SSL effectiveness in the radio astronomy domain.
\item Investigate the transferability of self-supervised representations across different domain-specific radio astronomy datasets.
\item Provide insights into the most effective SSL techniques for radio astronomical data analysis.
\end{itemize}
We conduct experiments using a range of state-of-the-art SSL methods, including SimCLR \cite{chen2020simplesimclr}, BYOL \cite{grill2020byol}, DINO \cite{caron2021dino}, WMSE \cite{ermolov2021whitening}, SwAV \cite{caron2020swav} and All4One\cite {estepa2023all4one}. These methods are applied to both curated and uncurated radio astronomy datasets. Our evaluation encompasses multiple downstream tasks, focusing on source classification across diverse datasets such as Radio Galaxy Zoo (RGZ) \cite{banfield2015radiorgz}, MiraBest \cite{mirabest}, and VLASS \cite{gordon2023vlass}. 
Additionally, we present the Multi-Survey Radio Sources (MSRS) dataset, a curated collection from four existing radio surveys, labeled according to a new schema specifically developed for this study. This dataset provides a unique resource for evaluating self-supervised learning methods across different radio surveys and source morphologies.

Our results demonstrate the potential of SSL in radio astronomy, consistently outperforming ImageNet pre-trained baselines across all datasets, highlighting the value of domain-specific pre-training even by simply performing linear adaptation of SSL features.
By providing this comprehensive benchmark of SSL methods in radio astronomy, we aim to contribute to the development of effective and efficient techniques for leveraging the vast amounts of unlabeled data in this domain. These insights may prove valuable not only for upcoming large-scale projects like SKA but also for informing similar approaches in other scientific fields characterized by abundant unlabeled data and domain-specific challenges.
\section{Related works}
Computer vision techniques have become increasingly important in astronomy, finding applications across various wavelengths, including infrared, optical, and radio. Traditionally, machine learning approaches in astronomy have focused on unsupervised learning methods to extract representations from astronomical images, which are then visually explored using dimensionality reduction algorithms. These feature extraction techniques include autoencoders \cite{bordiu2022patterns}, self-organizing maps (SOM) \cite{mostert2021unveilingsom}, and SSL \cite{sarmiento2021capturing,mohale2024enablingssldiscovery}.
The extracted representations serve multiple purposes beyond visual inspection, including anomaly detection, classification, and instance segmentation. 

In addressing these tasks, radio astronomy has followed a logical progression mirroring the broader evolution of computer vision techniques. Initially, astronomers primarily relied on supervised learning methods \cite{riggi2024classification}, favoring this approach due to its historical precedence and relative simplicity in implementation and interpretation.
As the field advanced, researchers began to explore more sophisticated techniques, leading to the adoption of SSL in astronomy. However, a recent survey \cite{huertas2023brief} notes that while SSL methods have gained traction, they have been applied primarily to non-radio images, such as optical data. This highlights a gap in the application of these techniques to radio astronomy, which presents unique challenges and opportunities.

In the specific domain of radio astronomy, recent work by Slijepcevic et al. \cite{slijepcevic2023foundational} demonstrates the potential of SSL methods. They employed BYOL~\cite{grill2020byol} on the Radio Galaxy Zoo Data Release 1 (RGZ-DR1) dataset~\cite{banfield2015radiorgz} to pretrain a general model applicable to various downstream tasks. The model's performance was quantitatively evaluated using the MiraBest dataset~\cite{mirabest}, which provides physically meaningful morphological classifications. However, this evaluation was limited by the relatively small size of the MiraBest dataset (about 800 images) and its binary classification schema (FRI vs. FRII radio galaxies).
Riggi et al. \cite{riggi2024self} addressed this limitation by constructing both curated and uncurated unlabeled datasets, reserving the labeled RGZ-DR1 data for the evaluation phase.
\section{Materials and Methods}

\subsection{Overview}

Deep radio sky observations are nowadays carried out with large arrays of radio telescopes, that collect sky visibility data across multiple frequency channels. These raw data undergo complex interferometric processing, including calibration and imaging, to produce either single-frequency radio continuum maps or multi-frequency spectral-line data cubes. Our study focuses on radio-continuum maps, which are single-channel grayscale images in FITS format. These images represent radio flux brightness in Jy/beam, with pixel values ranging from $\mu$Jy/beam to several Jy/beam, including negative values often associated with imaging artifacts. The radio continuum maps generated by each survey presents different resolutions, but generally consist of very large images (e.g., for SMGPS \cite{goedhart2023sarao}, 7500$\times$7500 pixels).

In this section, we present the methodology carried out for our systematic study of SSL approaches for radio-astronomy data analysis. We first introduce the dataset employed in this work for pretraining backbone models in a self-supervised fashion and describe data preprocessing modalities. We then introduce the variety of SSL techniques employed in this work, briefly presenting their characteristics and training objectives. Finally, we present the list of publicly-available datasets used in this work as downstream tasks, and the evaluation procedure for assessing the performance of SSL pretraining on those tasks.

\subsection{Self-supervision datasets}
Compared to traditional supervised learning, SSL approaches provide the important advantage of not requiring manual labeling of data samples, which is well-known as a time-consuming and error-prone task. However, while natural image datasets are inherently built ensuring that each data sample has meaningful and somewhat unique content, radio-astronomy data present significant challenges in this regard. 

The easiest way to build a large \emph{uncurated} radio dataset is to randomly extract (i.e., without any knowledge of the position of radio sources) cutout images from radio maps using a sliding window with fixed-size. This procedure can potentially sample a high variety of object morphologies, but, as the sky is dominated by compact point-like sources and by background, while the number of peculiar and extended objects is significantly smaller, the result is the unavoidable construction of an unbalanced dataset, where more interesting objects (e.g., diffused or extended sources) are relatively rare. As is known \cite{assran2022unbalance}, SSL methods suffer when dealing with unbalanced data. Additionally, considering the multi-scale nature of objects in the sky, using a fixed size of the sliding window likely results in truncated or partially captured sources.

An alternative approach consists in building a \emph{curated} dataset by extracting cutouts around known source celestial positions reported in existing radio source catalogues. In this case, it is possible to adaptively set the image cutout size to be large enough to fully include the catalogued source and part of its surrounding region (including the background or other nearby sources). As can be imagined, source catalogues need to be manually labeled, and inevitably include fewer objects than the totality that can be found in radio maps.

In our work, we assess the impact of data curation by using two different datasets for SSL training, indicated in the following as \emph{Curated} and \emph{Uncurated} dataset. These datasets are primarily collected from two radio surveys:

The SARAO MeerKAT Galactic Plane Survey (SMGPS)~\cite{goedhart2023sarao}: Covers a large portion of the 1st, 3rd and 4th Galactic quadrants ($l=2^{\circ}$-$61^{\circ}$, $251^{\circ}$-$358^{\circ}$, $|b|<1.5^{\circ}$) in the L-band (886-1678 MHz), with 8" angular resolution and $\sim$10-20 $\mu$Jy/beam noise rms at 1.3 GHz.

The ASKAP EMU pilot survey~\cite{norris2011emu}: Covers approximately 270 deg$^2$ of the Dark Energy Survey area, with 11"-18" angular resolution and $\sim$30 $\mu$Jy/beam noise rms at 944 MHz.

\begin{figure}
\includegraphics[width=1.\textwidth]{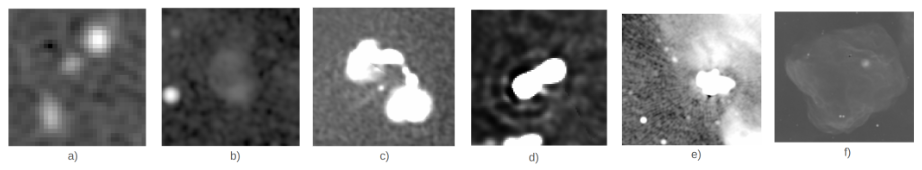}
\caption{Different visual characteristics of radio images. a) a multi-island radio source in low resolution; b) a faint diffuse source enhanced through a log scale transform; c) \textit{mosaicking} artifact shown as a diagonal step line; d) \textit{water ripple} artefact pattern around a bright source e) a large-scale diffuse emission region; f) a very large diffuse source with various nested compact sources along the line of sight.
} \label{artefacts}
\end{figure}

\noindent{\textbf{Uncurated dataset.}} A set of 285,585 radio images of fixed size (256$\times$256 pixels, equivalent to a $\sim$6.4'$\times$6.4' sky portion). As stated before, data are extracted from radio maps using a sliding windows, with a 50\% overlap. Since an intrinsic limitation of that dataset is the fixed sliding window size, we choose it to be large enough to capture most of the extended sources in the maps. As the data are extracted from mosaicked maps, those images may contain missing values on the border (filled in with the minimum value from the corresponding cutout) and mosaicking artifacts (see Fig.~\ref{artefacts}).

\noindent\textbf{Curated dataset.} A collection of 17,062 radio images, derived from the SMGPS integrated maps. These images are centered on objects cataloged in the SMGPS extended source catalogue \cite{goedhart2024sarao}. 
Unlike fixed-size datasets, images have variable dimensions, each scaled to 2.5 times the bounding box of its central object. This adaptive sizing ensures comprehensive capture of source structures. The dataset encompasses a rich variety of radio source morphologies, including multi-component sources (e.g. radio galaxies), and diffuse structures. 

\begin{figure}
    \centering
    \begin{subfigure}[b]{0.48\textwidth}
        \centering
        \includegraphics[width=0.32\textwidth]{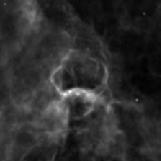}
        \includegraphics[width=0.32\textwidth]{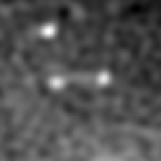}
        \includegraphics[width=0.32\textwidth]{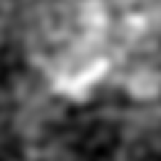}
        \caption{Curated}
    \end{subfigure}%
    \hfill
    \begin{subfigure}[b]{0.48\textwidth}
        \includegraphics[width=0.32\textwidth]{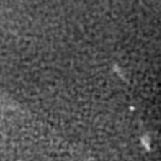}
        \includegraphics[width=0.32\textwidth]{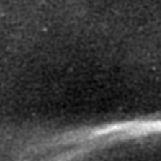}
        \includegraphics[width=0.32\textwidth]{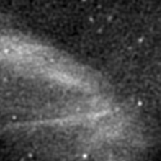}
        \caption{Uncurated}
        
    \end{subfigure}%
    \caption{Images extracted from Curated and Uncurated dataset. Curated samples correspond to well-fit crops of radio sources, while uncurated ones generally include more background and uncentered or partially-cropped objects.}
    \label{fig:pretrain-examples}
    \hfill
\end{figure}

Examples of images extracted from the Curated and Uncurated datasets are presented in Fig.~\ref{fig:pretrain-examples}.

\subsection{Self-supervision methods}
From the plethora of methods available from the state of the art, we select a subset of SSL techniques minimizing the overlap within training strategies, to provide readers with a comprehensive analysis. Given the limited research on applying SSL methods to radio-astronomy images and in order to favor a comparison with the literature, we believe it is prudent to build a solid baseline with well-established CNN models, leaving out vision transformers (as they require significantly amount of computational resources and since those can exhibit instability during training). Therefore, methods that principally rely on ViT~\cite{Dosovitskiy2021vit} (e.g. MAE~\cite{he2022mae} or DinoV2~\cite{oquab2023dinov2}) are not considered.

Additionally, we focus on methods based on view augmentation rather than on pretext tasks, since the latter may not make sense with some kinds of radio sources: for instance, some sources may be rotation-invariant, while the large amount of background in certain images (especially in the uncurated dataset) hinders the application of inpainting/jigsaw-based tasks. 

In the following, we present an overview of the SSL methods employed in this study. As mentioned above, the methods under analysis all involve the generation of two augmented views, $x'$ and $x''$, from the same starting image $x$, by means of random method-specific transformations. This approach is pivotal in learning robust feature representations, as it enables the model to understand and capture the intrinsic properties of the images across possible variants.

In \textbf{SimCLR}~\cite{chen2020simplesimclr}, the views are processed by a model producing representations $z'$ and $z''$. The method relies on attracting representations of views generated by the same image, while repelling views generated by different images. To this aim, SimCLR uses a projection network and a loss defined as:
\[
     \mathcal{L}_\text{SimCLR} = -\log \frac{\exp(\text{sim}(h_i, h_j) / \tau)}{\sum_{k=1}^{2N} \mathbf{1}_{[k \neq i]} \exp(\text{sim}(h_i, h_k) / \tau)}
     \]
where \( h_i \) and \( h_j \) are the projections of $z'$ and $z''$,   \(\text{sim}\) is cosine similarity, and \(\tau\)  is a temperature parameter.

\textbf{BYOL}~\cite{grill2020byol} tackles the problem from a slightly different perspective, without leveraging negative examples. It involves two networks, \emph{online} and \emph{target}, and a predictor on top of the online projector. Both networks are trained simultaneously in a teacher-student fashion, with the online target attempting to predict the target's representations; in turn, the target network does not receive parameter updates through gradient descent, but its parameters are obtained through an exponential moving average of the student's. BYOL's loss can be summarized as:
\[
     \mathcal{L}_\text{BYOL} = \|q_\theta(z_{\text{online}}) - z_{\text{target}}\|_2^2
     \]
where \( q_\theta \) is the predictor network, and $z_\text{online}$ and $z_\text{target}$ are the representations obtained by the online and target networks, respectively.

\textbf{DINO}~\cite{caron2021dino} addresses SSL using a similar teacher-student setting in a knowledge distillation framework, with the student network predicting the output of the teacher with a standard cross-entropy loss:
\[
     \mathcal{L}_\text{DINO} = H(\sigma(z_t / \tau_t), \sigma(z_s / \tau_s))
     \]
where \(z_t\) and \(z_s\) are the outputs of the teacher and student networks, respectively, \textit{\textbf{\(\sigma\)}} denotes the softmax function, and \(\tau_t\) and \(\tau_s\) are temperature parameters.

\textbf{WMSE}~\cite{ermolov2021whitening} employs a single encoder network and positive samples only, preventing feature collapse by using a whitening operation that maps the representation space into a zero-mean and identity-covariance distribution.
The loss could be represented as: uses the mutual information maximization in combination with whitening the representations.
\[
     \mathcal{L}_\text{WMSE} =  \| W(z') - W(z'') \|_2^2
     \]
where  \({W}(z)\) denotes the whitening transformation applied to representation \( z \).

Clustering is traditionally one of the most suitable methods for unsupervised analysis. \textbf{SwAV}~\cite{caron2020swav} adapts clustering to SSL by assigning pseudo-labels to different views of the same image.
Given views $x'$ and $x''$ of the same image, SwAV trains a model to compute features $z'$ and $z''$, which are then mapped to soft assignments $q'$ and $q''$ based on their similarity to a set of prototypes $C$. Then, the model is trained to predict the soft assignment of one view from the representation of the other view:
\[
     \mathcal{L}_{\text{SwAV}} = \ell(q', z'') + \ell(q'', z')
\]
where ($\ell$) is the cross-entropy.

Other approaches, such as NNCLR~\cite{dwibedi2021nnclr}, propose to increase the diversity of positive pairs by pulling together a view of a sample with the nearest neighbor (NN) among the augmented views of another sample. \textbf{All4One}~\cite{estepa2023all4one} builds upon this concept and extends it by efficiently including multiple neighbors through a self-attention mechanism and integrating a redundandy reduction loss inspired by Barlow Twins~\cite{zbontar2021barlow}.

In our experiments, we use the implementations of the above methods provided by the \emph{solo-learn}~\cite{sololearn}, ensuring that all experiments are implemented with a consistent standard, reducing variability and potential biases that might arise from different coding practices.

\subsection{Downstream datasets}
\begin{figure}
    \centering
    
    \begin{subfigure}[b]{0.48\textwidth}
        \centering
        \includegraphics[width=0.32\textwidth]{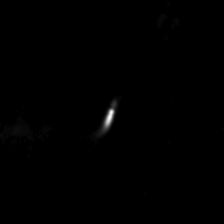}
        \includegraphics[width=0.32\textwidth]{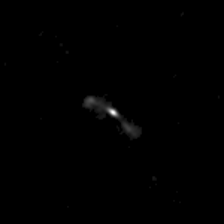}
        \includegraphics[width=0.32\textwidth]{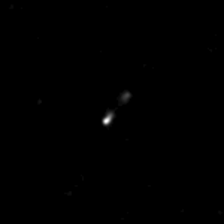}
        \caption{MiraBest}
    \end{subfigure}%
    \hfill
    \begin{subfigure}[b]{0.48\textwidth}
        \centering
        \includegraphics[width=0.32\textwidth]{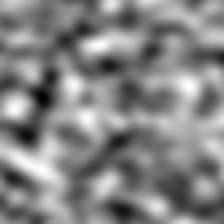}
        \includegraphics[width=0.32\textwidth]{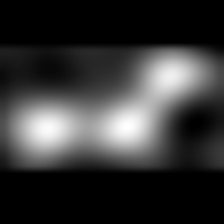}
        \includegraphics[width=0.32\textwidth]{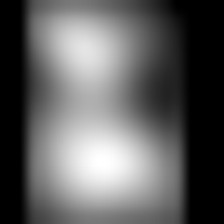}
        \caption{MSRS}
    \end{subfigure}%
    \hfill
    
    \vspace{1em}

    \begin{subfigure}[b]{0.48\textwidth}
        \centering
        \includegraphics[width=0.32\textwidth]{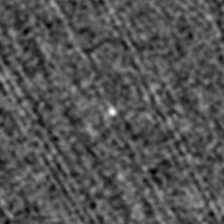}
        \includegraphics[width=0.32\textwidth]{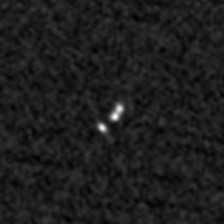}
        \includegraphics[width=0.32\textwidth]{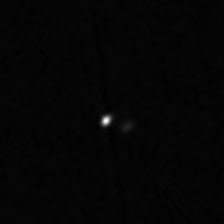}
        \caption{VLASS}
    \end{subfigure}%
    \hfill
    \begin{subfigure}[b]{0.48\textwidth}
        \centering
        \includegraphics[width=0.32\textwidth]{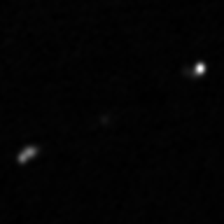}
        \includegraphics[width=0.32\textwidth]{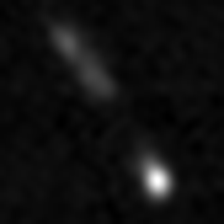}
        \includegraphics[width=0.32\textwidth]{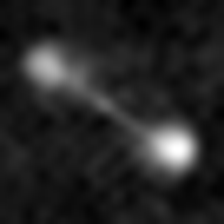}
        \caption{RGZ}
    \end{subfigure}%
    \hfill
    \caption{Image samples for the downstream datasets employed in our study.}
    \label{fig:downstream-examples}
\end{figure}

To assess the effectiveness of self-supervised pretraining across the methods under analysis, we utilize publicly available radio-astronomy classification benchmarks as \emph{downstream tasks}.
We take into account datasets generated from various sky surveys, each encompassing distinct source types. Each dataset exhibits unique visual characteristics, as shown in Fig.~\ref{fig:downstream-examples}. It should be noted that the original versions of the employed datasets feature a large class imbalance. Since this work addresses the quality of SSL representations, we resample each dataset so that all classes are balanced, either by undersampling more populated classes or by duplicating samples from less populated ones. The total number of samples included in each dataset after resampling is reported in the following.

\noindent\textbf{Multi Survey Radio Sources (MSRS).} This dataset is a collection of sources of different morphologies observed in various radio surveys (FIRST~\cite{becker1995first}, EMU~\cite{norris2011emu}, SCORPIO~\cite{umana2015scorpio}, SMGPS~\cite{goedhart2024sarao}), covering galactic and extragalactic plane regions and showing different SNR ratios, angular resolutions, artifact patterns.
Sources were labelled according to the following taxonomy:
\emph{1C-NP}: small single-island sources with N peaks, e.g., point-like (N=1), double (N=2), triple (N=3);
\emph{Diffuse}: faint diffuse structures with roundish or irregular shape;
\emph{Extended}: single-component sources with extended morphology;
\emph{Extended-MI}: Multi-island extended sources, consisting in disjoint regions belonging to the same source.
Besides being a multi-survey dataset, this is the only downstream dataset considered in this work that include samples of diffuse sources (the most challenging class) and images with pure background noise. Image cutouts are rectangular and equal to the original source size. The total number of samples in this dataset is 11,550.

\noindent\textbf{Radio Galaxy Zoo (RGZ)}~\cite{banfield2015radiorgz} is retrieved from the crowd labeling campaing on Zooniverse\footnote{https://www.zooniverse.org/}. This includes radio images from the VLA Faint Images of the Radio Sky at Twenty cm (FIRST) extragalactic survey (1.4 GHz, angular resolution $\sim$5") \cite{becker1995first}. We use the data release 1, where angular size is also available for each source, therefore giving us the abilty to suitably crop the image around the source, extracting squared bounding boxes with side equal to 1.5 times the source size. The dataset classification schema includes 6 classes comprising different amount of components C and peaks P, namely: 1C-1P, 2C-1P, 2C-2P, 3C-1P, 3C-2P, 3C-3P. The resulting dataset includes 27,000 samples.

\noindent\textbf{MiraBest}~\cite{mirabest} is a small dataset comprising FRI and FRII radio galaxies, as well as hybrid sources from extragalactic plane regions. For comparison with \cite{slijepcevic2023foundational} we consider the sources tagged as ``certain'' and discarded hybrid source. Cutout size is fixed to 150$\times$150 pixels. The dataset contains 397 FRI samples and 435 FRII samples, for a total of 832 (we do not perform resampling in this case).

\noindent\textbf{VLASS} is a survey~\cite{gordon2023vlass} covering galactic and extragalactic plane regions. We use Quick Look epoch 1 version 3 and extract sources from the Table 2 of the catalogue\footnote{https://cirada.ca/vlasscatalogueql0}, providing radio loud sources associated to their host spotted in the infrared band. The original source cutouts have a 500$\times$500 size, probably to include the host galaxy in the infrared band, which however leads to the inclusion of a lot of background. For this reason, we reduce the cutout to 224$\times$224: the background is still wide, but reasonable. The taxonomy of sources within the dataset includes: single-component sources; sources with two close components; sources with three close components; sources with two asymmetric radio components, many of which may be instances of a radio core blended with a lobe; sources that are notably brighter than their close neighboring components in the radio frequency. The total size of the resampled dataset is  14,500.

\subsection{Downstream evaluation}
Following the literature on SSL, we compare the performance of the methods under analysis by carrying out a \emph{linear evaluation} on the downstream tasks, i.e., by directly training a linear classifier mapping output features from the SSL backbone to the target classes. This procedure is intended to directly measure whether the representation learned by the model contains distinguishing features for the target classes. Additionally, given the relatively small size of the target datasets, we also perform \emph{fine-tuning} of the SSL backbone on the downstream tasks, to investigate the effect of directly updating backbone features.

\section{Experimental results}

\subsection{Training and evaluation details}
Following common practice in radio-astronomy, input images are normalized using the minimum and maximum values within a single cutout; we then resize them to 224$\times$224. We employ both ResNet-18 and ResNet-50 as backbones for SSL. All methods are trained for using the LARS~\cite{chowdhury2021evaluatinglars} optimizer, with a batch size of 512. Training on the Uncurated dataset is carried out for 100 epochs; on the Curated dataset, since it is significantly smaller, we train for 600 epochs. For all augmentation-based SSL methods, we apply the following set of transformations, with a certain probability $p$: horizontal/vertical flip ($p = 0.5$); Gaussian blur with $\sigma$ between 0.1 and 2 ($p=0.25$); contrast adjustment by a random value between 0.2 and 1.8 ($p=0.5$); random crop with scale between 0.65 and 1 ($p=1$). The selection of other hyperparameters is carried out independently for each SSL method, by manually varying key parameters and observing the average loss on the Curated and Uncurated datasets. In the following, we detail the final hyperparameters chosen for each method:
\begin{itemize}
\item \textbf{SimCLR}. Base learning rate: 1.2; output projection size: 512; temperature: 0.2.
\item \textbf{BYOL}. Base learning rate: 1.2; projection size: 512; predictor hidden size: 1024.
\item \textbf{DINO}. Base learning rate: 0.016; projection size: 256.
\item \textbf{WMSE}. Base learning rate: 0.002; projection size: 128; whitening size: 256.
\item \textbf{SWAV}. Base learning rate: 1.2; projection size: 128; number of prototypes: 300; temperatur: 0.1.
\item \textbf{All4One}. Base learning rate: 1.0; projection size: 512; predictor hidden size: 4096; temperature: 0.2.
\end{itemize}

When training on a downstream task with the fine-tuning strategy, we employ the AdamW~\cite{loshchilov2017decoupled} optimizer with a batch size of 256 and a learning rate of 0.0005, with a linear warmup followed by a cosine annealing schedule. For linear evaluation, we use a standard SGD optimizer, with the same batch size and initial learning rate. We employ a step scheduler, with learning rate decay steps of 0.1 factor at epochs 10 and 80. For both fine-tuning and linear evaluation, the total number of epochs is 100. During downstream training, we apply random vertical/horiziontal flip ($p=0.5$) and random crop with scale between 0.95 and 1 ($p=1$).

Evaluation results on the downstream tasks are reported in terms of classification accuracy. Using the above final hyperparameters, we train each SSL method with each backbone on each pretraining dataset for three times with random initialization. The only exception is that, for the Uncurated dataset, we only use ResNet-18, for timing constraints. Then, we evaluate each trained model on all downstream tasks, using 3-fold cross-validation on each task. In practice, for a given combination of SSL method, backbone, pretraining dataset and downstream dataset, we have nine values of accuracy, for which we report the corresponding mean and standard deviation.

As an additional baseline for comparison, we also report the results obtained when pretraining BYOL on the ImageNet-100 (for ResNet-18) and ImageNet-1k (for ResNet-50) datasets. This provides useful information on the suitability of features extracted from natural images when applied to the analysis of radio-astronomy imaging data.

All experiments we carried out on a single NVIDIA A100-PCIE-40GB GPU.

\subsection{Linear evaluation}

\begin{table}[ht]
    \caption{Linear evaluation: mean accuracies and standard deviations of each configuration. Best results in bold for each block.}
    \label{tab:linear_minmax_results}
    \centering
    \resizebox{0.95\textwidth}{!}
    {%
    \begin{tabular}{lll|c|c|c|c}
    \toprule
        \textbf{Dataset} & \textbf{Backbone} & ~~\textbf{Method~~} & \textbf{MiraBest} & \textbf{RGZ} & \textbf{MSRS} & \textbf{VLASS} \\
        \toprule
        \multirow{6}{*}{Curated} & \multirow{6}{*}{ResNet-18} & ~~All4one    &  82.1 ± 0.5            & \textbf{79.4 ± 0.2}   & \textbf{78.2 ± 3.9}   & \textbf{77.2 ± 0.8} \\
         &  & ~~BYOL       &  89.6 ± 0.4            & 77.6 ± 0.2            & 78.0 ± 4.2   & 76.2 ± 0.6 \\
         &  & ~~DINO       &  64.2 ± 0.4            & 69.2 ± 0.6            & 73.8 ± 3.8            & 66.6 ± 0.8 \\
         &  & ~~SimCLR     &  \textbf{91.0 ± 0.5}   & 69.5 ± 0.5            & 73.7 ± 4.1            & 71.9 ± 1.1 \\
         &  & ~~SwAV       &  72.9 ± 0.4            & 74.6 ± 0.7            & 74.8 ± 2.9            & 69.6 ± 1.3 \\
         &  & ~~WMSE       &  84.6 ± 0.0            & 70.6 ± 0.5            & 74.6 ± 4.6            & 70.6 ± 0.0 \\
        \midrule                                    
        \multirow{6}{*}{Curated} & \multirow{6}{*}{ResNet-50} & ~~All4one    &  88.5 ± 0.6            & \textbf{78.8 ± 0.2}   & \textbf{77.1 ± 3.6}   & \textbf{77.0 ± 0.4} \\
         &  & ~~BYOL       &  \textbf{90.0 ± 0.5}   & 78.6 ± 0.5            & 76.5 ± 4.8            & 76.8 ± 0.4 \\
         &  & ~~DINO       &  77.7 ± 1.1            & 70.2 ± 0.6            & 73.6 ± 3.5            & 67.7 ± 1.4 \\
         &  & ~~SimCLR     &  85.8 ± 0.9            & 73.0 ± 0.1            & 71.7 ± 4.2            & 73.1 ± 0.9 \\
         &  & ~~SwAV       &  82.3 ± 0.5            & 75.3 ± 0.2            & 74.4 ± 2.3            & 70.5 ± 0.5 \\
         &  & ~~WMSE       &  81.2 ± 0.5            & 74.7 ± 0.3            & 75.7 ± 4.4            & 72.6 ± 0.4 \\
        \midrule                                    
        \multirow{6}{*}{Uncurated} & \multirow{6}{*}{ResNet-18} & ~~All4one  &  75.6 ± 0.8            & 68.4 ± 0.5            & \textbf{74.2 ± 3.8}   & \textbf{70.5 ± 0.4} \\
         &  & ~~BYOL     &  79.6 ± 0.4            & \textbf{72.6 ± 0.9}   & 73.5 ± 3.0            & 69.6 ± 0.2 \\
         &  & ~~DINO     &  83.6 ± 0.0            & 67.4 ± 1.0            & 71.4 ± 3.6            & 69.0 ± 0.8 \\
         &  & ~~SimCLR   &  \textbf{84.8 ± 0.7}   & 68.2 ± 0.3            & 72.0 ± 3.4            & 68.1 ± 0.8 \\
         &  & ~~SwAV     &  74.4 ± 0.8            & 65.2 ± 0.9            & 73.2 ± 3.1            & 62.5 ± 0.6 \\
         &  & ~~WMSE     &  64.6 ± 0.4            & 60.2 ± 0.5            & 69.6 ± 3.5            & 60.5 ± 0.6 \\
        \midrule                                    
        ImageNet-100~~ & ResNet-18 & ~~BYOL   &  67.5 ± 0.9            & 63.6 ± 0.5            & 72.0 ± 4.2            & 62.8 ± 0.9 \\
        ImageNet-1k    & ResNet-50 & ~~BYOL   &  73.5 ± 1.0            & 70.2 ± 0.8            & 76.0 ± 4.9            & 69.9 ± 0.3 \\
        \bottomrule
    \end{tabular}
    }
\end{table}

Results for linear evaluation are reported in Table~\ref{tab:linear_minmax_results}. A high-level analysis across methods shows that All4one, BYOL and SimCLR generally achieve the best performance on the downstream tasks, while DINO, SwAV and WMSE seem to perform worse on average. In particular, All4one yields the highest accuracy on three downstream tasks out of four, excluding MiraBest. The superior trend of All4one is confirmed when varying across the backbone architectures, as well as on both the Curated and Uncurated pretraining datasets.

From a quantitative perspective, dataset curation positively impacts results, as all SSL methods benefit from the higher sample quality more than from a larger dimension of the dataset. Interestingly, even though MSRS partially overlaps with the Uncurated pretraining dataset (since both contain some of the same sky regions) the models pretrained on the Curated dataset perform better. This is likely because the Curated dataset includes entire, more complex structures, whereas the Uncurated dataset contains only portions of these structures. As a result, pretraining on the Curated dataset allows the models to learn more comprehensive and transferable features. Backbone architecture has a more limited effect, with ResNet-18 generally yielding slightly better results than ResNet-50, which can be easily explained by the simplicity of the image patterns, not requiring a particularly high architectural complexity.

It is interesting to note that the SSL baseline using ImageNet variants almost always performs significantly worse than when using radio-astronomy data for pretraining. Only in the case of MSRS, which is characterized by larger and more structured object shapes, do the baselines yield closer (but still lower) accuracy, which might indicate that features learned from natural images may be overly complex (and thus less transferable) for the tasks at hand. However, it should also be noted that MSRS exhibits a significantly higher standard deviation, compared to the other downstream datasets. Hence, the similarity in terms of accuracy may also be due to an instability in the representations learned during self-supervision. Further investigations are therefore in order to clarify this aspect.

\begin{table}[ht]
\caption{Fine-tuning: mean accuracies and standard deviations of each configuration. Best results in bold for each block.}
    \label{tab:finetune_minmax_results}
    \centering
    \resizebox{0.95\textwidth}{!}{%
    \begin{tabular}{lll|c|c|c|c}
    \toprule
        \textbf{Dataset} & \textbf{Backbone~~} & \textbf{Method} & \textbf{MiraBest} & \textbf{RGZ} & \textbf{MSRS} & \textbf{VLASS} \\
        \toprule
        \multirow{6}{*}{Curated}      & \multirow{6}{*}{ResNet-18} & All4one    & 96.2 ± 0.6            & 81.3 ± 0.5            & 76.7 ± 4.4            & 82.9 ± 0.3 \\
              &  & BYOL       & 96.5 ± 1.0            & \textbf{81.6 ± 0.2}   & 76.5 ± 5.0            & \textbf{83.5 ± 0.2} \\
              &  & DINO       & 97.1 ± 0.6            & 80.0 ± 0.2            & 75.1 ± 4.1            & 82.1 ± 0.4 \\
              &  & SimCLR     & 94.2 ± 1.4            & 80.6 ± 0.3            & 76.0 ± 4.4            & 83.4 ± 0.6 \\
              &  & SwAV       & 94.2 ± 1.0            & 78.8 ± 0.4            & 76.8 ± 3.8            & 78.6 ± 1.3 \\
              &  & WMSE       & \textbf{99.2 ± 0.4}   & 81.1 ± 0.6            & \textbf{77.0 ± 4.1}   & 81.1 ± 1.0 \\
        \midrule
        \multirow{6}{*}{Curated}      & \multirow{6}{*}{ResNet-50} & All4one    & 95.0 ± 1.4            & 82.1 ± 0.3            & 76.7 ± 4.6            & 84.4 ± 0.2 \\
              & & BYOL       & 95.8 ± 1.8            & \textbf{82.6 ± 0.4}   & 75.4 ± 4.1            & 84.9 ± 0.2 \\
              &  & DINO       & \textbf{98.1 ± 0.9}   & 81.0 ± 0.4            & 74.8 ± 3.3            & \textbf{85.1 ± 1.1} \\
              &  & SimCLR     & 95.0 ± 1.4            & 82.1 ± 0.2            & 76.7 ± 3.6            & 84.2 ± 0.6 \\
              &  & SwAV       & 92.1 ± 0.7            & 80.9 ± 0.1            & 76.7 ± 3.9            & 81.7 ± 0.4 \\
              &  & WMSE       & 93.8 ± 1.2            & 81.5 ± 0.2            & \textbf{76.8 ± 3.6}   & 83.8 ± 0.6 \\
        \midrule
        \multirow{6}{*}{Uncurated}        & \multirow{6}{*}{ResNet-18} & All4one    & \textbf{96.5 ± 0.5}   & 80.7 ± 0.4            & 75.2 ± 3.6            & 83.1 ± 0.4 \\
                &  & BYOL       & 94.8 ± 1.6            & \textbf{81.0 ± 0.2}   & \textbf{75.6 ± 4.5}   & \textbf{83.7 ± 0.3} \\
                &  & DINO       & 95.0 ± 0.7            & 79.3 ± 0.5            & 74.4 ± 4.5            & 81.1 ± 0.5 \\
                &  & SimCLR     & 95.6 ± 1.2            & 79.1 ± 0.1            & 75.0 ± 3.2            & 81.7 ± 1.0 \\
                &  & SwAV       & 94.6 ± 1.3            & 79.7 ± 0.3            & 75.3 ± 4.0            & 82.6 ± 0.4 \\
                &  & WMSE       & 93.5 ± 1.5            & 79.1 ± 0.2            & 74.8 ± 4.7            & 83.0 ± 0.9 \\
        \midrule
        ImageNet-100~~ & ResNet-18 & BYOL       & 96.2 ± 0.6            & 81.0 ± 0.2            & 75.4 ± 5.3            & 84.1 ± 0.6 \\
        ImageNet-1k    & ResNet-50 & BYOL       & 98.5 ± 0.8   & 81.4 ± 0.2   & 76.1 ± 6.1   & 83.9 ± 0.9 \\
                \midrule
        RGZ & ResNet-18 & BYOL & 98.1 ± 0.3 & - & - & - \\
        \bottomrule
    \end{tabular}
    }
\end{table}

\subsection{Fine-tuning evaluation}
Fine-tuning results are reported in Table~\ref{tab:finetune_minmax_results}. As can be expected, results are higher than the linear evaluation setting, since the backbone models' features are explicitly updated for each downstream task. Of course, this comes with a higher training cost, as gradients for the entire backbone must be computed at training time. In this setting, the differences between SSL methods observed for linear evaluation are basically flattened: there is no marked superiority of one approach over the others. Even the ImageNet-based baselines achieve results on par with models pretrained on the radio-astronomy datasets.

In this setting, we introduce, as an additional baseline for comparison, the results of the work by Slijepcevic et al.~\cite{slijepcevic2023foundational}, where a ResNet-18 is pretrained on RGZ through BYOL, and then fine-tuned on MiraBest. To the best of our knowledge, this study is the most similar to ours from the literature, although it is significantly more limited in scope. Also in this case, the results are in line with the ones obtained in our study: however, due to the relative high performance that all approaches are able to achieve for MiraBest, we suggest that other downstream datasets might be more suitable for benchmarking in future works.

Despite the lack of significant differences in fine-tuning results, it is important to note that some tasks require data representations that are agnostic to specific classification schemas. For instance, visual data exploration tasks using dimensionality reduction techniques benefit from more general representations. In these scenarios, non-finetuned models can still provide valuable insights, offering representations that are useful for exploratory data analysis rather than specific classification tasks.

\section{Conclusions}

In this work, we investigated the potential of self-supervised learning (SSL) for enhancing the analysis of radio astronomical data, notably outperforming traditional models pretrained on natural images in several domain-specific downstream tasks. Our results indicate that SSL-trained models, particularly those using the All4one method, achieve notable improvements in accuracy during linear evaluation, suggesting that SSL can effectively leverage the unique characteristics of radio interferometry images. Advantages of SSL become less pronounced in the fine-tuning setting, though they still surpass the performance of models pretrained on natural images. Another key finding is the importance of data curation, which positively impacts SSL performance more significantly than the sheer size of the dataset. 

Given that ResNet-50 did not outperform ResNet-18 (likely due to the simplicity of the image patterns) it might seem counterintuitive to explore more complex architectures like transformers. However, we propose that future work should investigate multimodal large language models and incorporate additional modalities such as infrared and optical bands. By integrating data from multiple spectral bands, these multimodal transformers could learn more meaningful and rich representations, potentially enhancing the analysis of radio astronomical data beyond what single-modality models can achieve.

\section{Acknowledgements}
This paper is supported by the Fondazione ICSC, Spoke 3 Astrophysics and Cosmos Observations. National Recovery and Resilience Plan (Piano Nazionale di Ripresa e Resilienza, PNRR)
Project ID CN 00000013 "Italian Research Center for HighPerformance Computing, Big Data and Quantum Computing"
funded by MUR Missione 4 Componente 2 Investimento 1.4:
Potenziamento strutture di ricerca e creazione di "campioni nazionali di R\&S (M4C2-19)" - Next Generation EU (NGEU). We also acknowledge partial funding from the INAF SCIARADA project.

We acknowledge the CINECA award under the ISCRA initiative, for the availability of high-performance computing resources and support. In particular, we sincerely thank Andrea Piltzer and Giuseppe Fiameni.

\bibliographystyle{splncs04}
\bibliography{bib}

\end{document}